\documentclass[12pt,pra,aps,amssymb,amsfonts,amsmath,tightenlines]{revtex4}

\usepackage[caption = false]{subfig}
\usepackage{graphicx}
\usepackage{dsfont} 

\newcommand{\half}{\mbox{$\textstyle \frac{1}{2}$}}
\newcommand{\re}{\mbox{$\rm e$}}
\newcommand{\ri}{\mbox{$\rm i$}}
\newcommand{\rd}{\mbox{$\rm d$}}

\usepackage{xcolor}
\usepackage{bm}
\usepackage{txfonts}
\usepackage{amsfonts}
\usepackage{braket}

\begin{document}

\title{Open quantum dynamics for plant motions}
\author{Dorje C. Brody$^{1,2}$}
\affiliation{$^1$Department of Mathematics, University of Surrey, 
Guildford GU2 7XH, UK} 
\affiliation{
\vspace{-0.3cm}
$^2$St Petersburg National Research University of Information 
Technologies, Mechanics and Optics, St Petersburg 197101, Russia
}

\vspace{0.2cm} 
\date{\today}

\begin{abstract}
\noindent 
Stochastic Schr\"odinger equations that govern the dynamics of open quantum systems are given by the equations for signal processing. In particular, the Brownian motion that drives the wave function of the system does not represent noise, but provides purely the arrival of new information. Thus the wave function is guided by the optimal signal detection about the conditions of the environments under noisy observations. This behaviour is similar to biological systems that detect environmental cues, process this information, and adapt to them optimally by minimising uncertainties about the conditions of their environments. It is postulated that information-processing capability is a fundamental law of nature, and hence that models describing open quantum systems can equally be applied to biological systems to model their dynamics. For illustration, simple stochastic models are considered to capture heliotropic and gravitropic motions of plants. The advantage of such dynamical models is that it allows for the quantification of information processed by the plants. By considering the consequence of information erasure, it is argued that biological systems can process environmental signals relatively close to the Landauer limit of computation, and that loss of information must lie at the heart of ageing in biological systems. 
\end{abstract}

\maketitle


\section*{Introduction}

Biological systems are constantly exposed to changing environments but 
manage to adapt to them by gathering information about the conditions of 
their surroundings and processing this information to arrive at best strategies. 
This ability to gather and process information about 
the surrounding environment does not require sophisticated organs like the 
brain of an animal. For instance, slime mould -- a large brainless amoeba-like 
cell -- can solve computationally difficult problems in a relatively short time 
\cite{slime_1,slime_2}. When the root of a plant encounters another root, it 
can detect whether it is the root of the same plant or it belongs to another 
plant \cite{GN}. 
The parasitic plant Cuscuta (dodder) are capable of processing olfactory 
signals to identify host locations and distinguish plants of differing nutrient 
levels \cite{Moraes}. 
Or, recent experiments suggest that common bean plants can 
detect the existence of objects in their vicinities without physical contacts 
\cite{Calvo}. These, and 
countless other examples, beg the questions how basic biological systems 
such as a single cell are capable of processing and storing information, 
and how can one model the dynamics of biological systems resulting from 
their information processing capabilities. 

The phenomenon of responding to the changing environment, however, 
is not restricted to biological systems. Take, for instance, a quantum 
system like an atom or even a single particle. When a 
quantum system is immersed in an 
open environment, its state responds to the conditions of the environment. 
The dynamics of the system can then be interpreted as 
the result of signal processing, for, as demonstrated below, 
the dynamical equation governing the wave function of the system (the 
stochastic unravelling of the Lindblad equation) is precisely the equation for 
the optimal signal detection. Hence a quantum system in an 
open environment behaves as if it is a micro information processor, just like an 
aggregate of quantum particles forming biological systems. From this point 
of view it is natural to enquire whether the dynamical equation for the 
evolution of the wave function in an open environment can be used to also 
describe behaviours of biological systems. 

The purpose of this paper is to explore the hypothesis that signal 
processing capability is fundamental to laws of nature, applicable 
to a wide range of systems including both quantum and biological, 
and examine its 
consequences. Evidently there are innumerable -- if not all -- biological 
phenomena that may be addressed from the viewpoint of signal processing. 
For definiteness we shall focus on the motions of plants, in 
particular, the phenomena known as 
heliotropism or phototropism -- the solar or light 
tracking of plants \cite{Harmer1}, and gravitropism -- 
the gravitational-field tracking of plant roots \cite{Sato}. 
While scientific 
study of heliotropism and gravitropism 
goes a long way \cite{Darwin,Bose,Yin}, the understanding 
of the biokinematics 
as well as the purposes for heliotropism and gravitropism 
are only emerging in recent years 
\cite{Harmer2,Goyal,Baluska,Kiss}. Our purpose here is 
to characterise qualitatively the dynamical behaviours of plants that result 
from information processing by employing familiar models used in open 
quantum systems. This not only demonstrates how models of signal 
detection are universally applicable to capture behaviours of quantum as 
well as biological systems, but also provides a new tool to model, and 
hence to make predictions about the statistics, of plant motions. Specifically,  
for a qualitative modelling of 
light-tracking motions of plants we shall employ a 
stochastic Schr\"odinger equation for which the Lindblad operator is 
given by the position operator \cite{Diosi1}, but for the 
Hamiltonian we take it to be simply the momentum operator. Borrowing 
techniques of signal processing, we derive an exact solution to the 
dynamical equation which describes the one-dimensional tracking of a 
deterministic motion. Similarly, for gravity-tracking motions 
of plant-root orientations we employ a stochastic Schr\"odinger equation 
where both the Lindblad operator and the Hamiltonian are given by the 
Pauli matrix, for which the exact solutions are known \cite{BH1}.

The significance of the introduction of dynamical models for plant motions 
under uncertain environments -- a concept hitherto missing in the literature 
-- is that such models allow for the quantification of the information processed, 
as well as information erased. From Landauer's principle 
\cite{LR}, then, erasure of information is accompanied by energy consumption 
and heat production. 
We believe that this information erasure process, resulting in the increase of 
entropy, must lie at the heart of ageing, or arrow of time, in biological systems. 
This point of view has been advocated in the past \cite{BCP,Trewavas}, 
though the 
introduction of concrete dynamical models for describing the behaviours of 
biological systems in response to the changes of their environmental conditions 
under the influence of noise has hitherto been missing. 
In particular, our model provides a rough estimate for the amount of 
energy consumption in plants due to erasure of processed information. 
For instance, 
if a daisy flower were to process information at the Landauer limit, 
then our model predicts 
that the flowers will consume of order $10^{-2}$ eV of energy overnight 
per each information-bearing cell by erasing the processed 
information to track the sun during the day.  
In the case of circumnutation of bean plants, our estimates 
show that the erasure 
of information about the location of a climbing support requires about 
$1$ eV of energy per each information-bearing cell at the Landauer limit. 
This indicates that inefficient information processing can be detrimental 
to their survival.
The fact that most plants produce little heat, in particular, supports our 
hypothesis that they are able to process 
information at a level significantly closer to the Landauer limit than mechanical 
devices at our disposal.

\section*{Signal processing quantum dynamics}

We begin by establishing the relation between the stochastic Schr\"odinger 
equation 
associated with Lindbladean dynamics and signal processing. This 
connection has long been envisaged by Belavkin and others \cite{Belavkin}, 
but was made explicit by Brody \& Hughston 
\cite{BH1,BH2,BH3} as an effective tool for solving, and 
thus arriving at weak solutions to certain types of stochastic 
Schr\"odinger equations. To start, consider the following 
problem in signal processing. We have an unknown quantity of interest, 
e.g., a signal or a message, represented by the random variable $L$. Let 
us assume that $L$ takes discrete values $\{l_i\}$ 
with the probabilities $\{p_i\}$. 
The true value of $L$ is unknown to the observer, 
who merely receives information about the value of the signal $L$ 
that is obscured by noise. Assuming that noise is additive and is 
modelled by a Brownian motion $\{B_t\}$, and that the signal is 
revealed in time at a constant rate $\sigma$, the noisy observation 
of the signal is characterised by the information process 
\begin{eqnarray} 
\xi_t = \sigma L t + B_t.
\end{eqnarray}
In the literature of signal detection, the signal-plus-noise 
time series $\{\xi_t\}$ such as the one defined here is called the observation 
process. 
However, because $\{\xi_t\}$ models the flow of information, while in the 
quantum context the term ``observation'' can have multiple meanings, we 
shall refer to $\{\xi_t\}$ as the information process.
Note that if there is a large number of random additive contributions to noise, 
then the law of large numbers implies that it is reasonable to assume that 
additive noise is normally distributed, making Brownian motion a reasonable 
candidate to model noise. 
The best estimate of $L$ that minimises the quadratic 
error, given the observed time series $\{\xi_s\}_{s\leq t}$ up to time $t$, is 
given by the conditional expectation 
\begin{eqnarray}
\langle L\rangle_t = \sum_{i} l_i \, {\mathbb P}\left(L=l_i|\{\xi_s\}_{s\leq t} 
\right) . 
\end{eqnarray} 
The problem thus reduces to working out the conditional probability 
$\pi_{it} = {\mathbb P}(L=l_i|\{\xi_s\}_{s\leq t})$ that $L=l_i$ given the 
observation. From the Markov property of the information process 
and the fact that $L=\lim_{t\to\infty}\xi_t/(\sigma t)$, the conditional 
probability reduces to a simpler expression $\pi_{it} = {\mathbb P}(L=l_i|\xi_t)$, 
which can easily be determined by use of the Bayes formula: 
\begin{eqnarray} 
{\mathbb P}(L=l_i|\xi_t) = \frac{{\mathbb P}(L=l_i)\,\rho(\xi_t|L=l_i)}
{\sum_j {\mathbb P}(L=l_j)\,\rho(\xi_t|L=l_j)} \, . 
\end{eqnarray}
Here $\rho(\xi_t|L=l_j)$ denotes the density function of $\xi_t$ conditional 
on the event that $L=l_j$. 
From ${\mathbb P}(L=l_i)=p_i$, and the fact that conditional on $L=l_i$ the 
random variable $\xi_t$ is normally distributed with mean $\sigma l_i t$ 
and variance $t$, we deduce that 
\begin{eqnarray}
\rho(\xi|L=l_i)=\frac{1}{\sqrt{2\pi t}} \exp\left(-
\frac{(\xi-\sigma l_i t)^2}{2t}\right), 
\end{eqnarray}
and hence, upon substitution, that 
\begin{eqnarray} 
\pi_{it} = \frac{p_i\,\exp\left(\sigma l_i\, \xi_t - \frac{1}{2} \sigma^2 l_i^2 t\right) }
{\sum_j p_j\,\exp\left(\sigma l_j\, \xi_t - \frac{1}{2} \sigma^2 l_j^2 t\right) } \, . 
\label{eq:3} 
\end{eqnarray}

We now consider taking the stochastic differential of the conditional 
probability process $\{\pi_{it}\}$. Because $\{\pi_{it}\}$ is a smooth function 
of the two variables $t,\xi_t$, by use of Ito’s formula 
\begin{eqnarray}
\rd f(t,\xi_t) = {\dot f}(t,\xi_t) \, \rd t + f'(t,\xi_t) \, \rd\xi_t + \half f''(t,\xi_t) \, \rd t 
\end{eqnarray} 
for any smooth function $f(t,x)$ of two variables, where dot denotes differentiation 
with respect to $t$ and dash denotes differentiation with respect to $x$, a 
calculation shows that 
\begin{eqnarray} 
\rd\pi_{it} = \sigma \pi_{it} \big( l_i - \langle L\rangle_t \big) \left( \rd\xi_t - \sigma 
\langle L\rangle_t \, \rd t \right).
\end{eqnarray}   
Introducing the process $\{W_t\}$ according to 
\begin{eqnarray}
W_t = \xi_t - \sigma \int_0^t \langle L\rangle_s \, \rd s   , 
\end{eqnarray} 
we can show that $\{W_t\}$ thus defined is a standard Brownian motion 
under the physical probability measure ${\mathbb P}$ 
\cite{BH2}. In signal detection, the process $\{W_t\}$ arising in this manner 
is called the innovations process \cite{Kailath}, and it reveals the arrival of 
new information. 
That is, the increment $\rd\xi_t$ in the observed time series contains both 
new information as well as redundant known information about the signal 
$L$. Removing $\sigma\langle L\rangle_t \,\rd t$ from this increment, we 
are left with only the new information that was not already known at time 
$t$. The fact that $\{W_t\}$ represents new information, however, follows 
only if $\langle L\rangle_t$ is the ``best'' estimate of $L$ 
that minimises the uncertainty (conditional variance) of 
$L$, exhausting all the 
relevant information gathered up to time $t$. We shall return to the 
significance of this statement in open quantum systems shortly, but for 
now we conclude that the increments $\rd\pi_{it} = \sigma\pi_{it}
( l_i - \langle L\rangle_t) \,\rd W_t$ of the conditional probability process 
are proportional to the increments of the innovations process. 

If we define $\phi_{it}=\sqrt{\pi_{it}}$ to be the square root probability, then 
another application of Ito's formula shows that 
\begin{eqnarray}
\rd\sqrt{\pi_{it}} = \frac{1}{2\sqrt{\pi_{it}}}\,\rd\pi_{it} - 
\frac{1}{8\pi_{it}\sqrt{\pi_{it}}}\,(\rd\pi_{it})^2 , 
\end{eqnarray} 
and hence from $(\rd\pi_{it})^2=\sigma^2 \pi_{it}^2 
(l_i-\langle L\rangle_t)^2\rd t$ that 
\begin{eqnarray}
\rd\phi_{it} = \textstyle{\frac{1}{2}} \sigma \big( l_i - \langle L\rangle_t \big) \, 
\phi_{it} \, \rd W_t - \textstyle{\frac{1}{8}} \sigma^2 
\big( l_i - \langle L\rangle_t \big)^2 \phi_{it} \, \rd t.
\label{eq:x9} 
 \end{eqnarray} 
With this expression in mind, let us consider a Hilbert space ${\mathcal H}$ 
associated 
with a physical system, on which an operator ${\hat L}$ is defined, whose 
eigenvalues are $\{l_i\}$ and eigenstates are $\{|l_i\rangle\}$. Assume that 
the Hamiltonian ${\hat H}$ of the system, whose eigenvalues are $\{E_i\}$,
commutes with ${\hat L}$, and that the eigenvalues of the two operators are 
not degenerate (the degenerate case can equally be treated \cite{BH2}). 
For simplicity let 
the initial state of the system be given by a pure state $|\psi_0\rangle$, and 
let $p_i=|\langle l_i|\psi_0\rangle|^2$. Let the process $|\psi_t\rangle$ be 
given by 
\begin{eqnarray}
|\psi_t\rangle = \sum_i \phi_{it} \, \re^{{\rm i}(\theta_i-E_i t)}\, | l_i\rangle, 
\label{eq:9} 
\end{eqnarray} 
where $\{\theta_i\}$ are arbitrary constant and $\phi_{it}$ satisfies 
(\ref{eq:x9}). 
Then from the discussion above it should be evident, by taking the stochastic 
differential of (\ref{eq:9}), that $|\psi_t\rangle$ \textit{is} 
the solution to the stochastic Schr\"odinger equation 
\begin{eqnarray} 
\rd|\psi_t\rangle =  - \ri {\hat H}\, |\psi_t\rangle \, \rd t + 
\textstyle{\frac{1}{2}} \sigma \big( {\hat L} - 
\langle {\hat L}\rangle_t \big) \, |\psi_t\rangle \, \rd W_t - \textstyle{\frac{1}{8}} 
\sigma^2 \big( {\hat L} - \langle {\hat L}\rangle_t \big)^2 |\psi_t\rangle \, \rd t , 
\label{eq:8} 
\end{eqnarray}
where 
\begin{eqnarray}
\langle {\hat L}\rangle_t = \langle\psi_t|{\hat L}|\psi_t\rangle = 
 \sum_{i} l_i \, {\mathbb P}\left(L=l_i|\xi_t\right) , 
\end{eqnarray} 
and the constants $\{\theta_i\}$ are fixed by the relative phases of the 
initial condition $|\psi_0\rangle$. This is 
the method introduced to find solutions to the so-called 
energy-based stochastic Schr\"odinger equation for which ${\hat L}={\hat H}$, 
that is, when the Lindblad operator is given by the 
Hamiltonian itself \cite{BH1}. But (\ref{eq:8}) is the dynamical equation for 
the wave function of 
the open quantum system that gives the stochastic unravelling of the 
corresponding Lindblad equation 
\begin{eqnarray} 
\partial_t {\hat\rho} = -\ri[{\hat H},{\hat \rho}] + \frac{1}{4}\sigma^2 
\left[ {\hat L} {\hat\rho} {\hat L} - \frac{1}{2} \left( 
{\hat L}^2 \rho + {\hat\rho} {\hat L}^2 \right) \right] 
\label{eq:z14} 
\end{eqnarray} 
for the reduced density matrix ${\hat\rho}$ of the system, so we are 
left with the signal-processing solution (\ref{eq:9}) 
to the stochastic unravelling dynamics, where $\phi_{it}=\sqrt{\pi_{it}}$ and 
where $\pi_{it}$ is given by (\ref{eq:3}). Putting it differently, if we define a 
matrix ${\hat\rho}$ by setting its $(i,j)$ element to be 
\begin{eqnarray}
{\hat\rho}_{ij} = \re^{-{\rm i}(E_i-E_j)t} \, {\mathbb E}\left[ \sqrt{\pi_{it}\,  
\pi_{jt}} \right] , 
\end{eqnarray} 
where ${\mathbb E}[-]$ denotes 
expectation over all random paths $\{\xi_t\}$, and if $\{\pi_{it}\}$ is the 
solution (\ref{eq:3}) to the signal detection problem, then the matrix 
${\hat\rho}$ satisfies the Lindblad equation (\ref{eq:z14}) for the 
dynamics of the open quantum system. In other words, 
the density matrix defined by ${\hat\rho}={\mathbb E}[
|\psi_t\rangle\langle\psi_t|]$ solves the Lindblad equation 
(\ref{eq:z14}) when the state vector $|\psi_t\rangle$ is the solution to 
(\ref{eq:8}). We remark that here we have restricted our discussion to 
Lindblad operators that are Hermitian, but an analogous conclusion 
can be deduced when they are not Hermitian, leading 
to the so-called quantum filtering equations \cite{Gough}. 

More generally, if $[{\hat H},{\hat L}]\neq0$, then we introduce a 
time-reversed state $|\varphi_t\rangle = \re^{{\rm i}{\hat H}t} |\psi_t\rangle$, 
and let ${\hat L}_t = \re^{{\rm i}{\hat H}t} \, {\hat L} \, \re^{-{\rm i}{\hat H}t}$. 
Then a short calculation shows that the stochastic Schr\"odinger equation 
for $|{\varphi_t}\rangle$ is given by 
\begin{eqnarray}
\rd |\varphi_t\rangle =  
\half \sigma ({\hat L}_t-\langle {\hat L}_t\rangle) 
|\varphi_t\rangle \, {\rd}W_t 
- \textstyle{\frac{1}{8}}  
\sigma^2({\hat L}_t-\langle {\hat L}_t\rangle)^2 |\varphi_t\rangle 
\, {\rd}t  , 
\label{eq:10} 
\end{eqnarray}
where $\langle {\hat L}_t\rangle = \langle{\varphi}_t| \,{\hat L}_t \, 
|\varphi_t\rangle$. Notice that in the time-reversed representation the 
unitary term associated with the Hamiltonian is removed from the 
dynamical equation. Thus, the solution for $|\varphi_t\rangle$ is obtained 
by finding the signal detection problem for which the time-dependent 
signal process $L(t)$ corresponds to the spectrum of the quantum operator 
$\re^{{\rm i}{\hat H}t} \, {\hat L} \, \re^{-{\rm i}{\hat H}t}$, and for the 
(non-Markovian) information process we have 
\begin{eqnarray} 
\xi_t = \sigma \int_0^t L(s)\,\rd s + B_t .
\end{eqnarray}  
Specifically, letting $\pi_{it}={\mathbb P}(L(t)=l_i|\{\xi_s\}_{s\leq t})$, the 
solution to (\ref{eq:10}) can be constructed by setting $|\varphi_t\rangle 
= \sum_i \sqrt{\pi_{it}} \, \re^{{\rm i}\theta_i}\, | l_i\rangle$.

It is worth remarking that a stochastic unravelling of the Lindblad equation 
for the reduced density matrix of the system in terms of a random pure-state 
dynamics is by no means unique, and thus is not restricted to the Ito 
stochastic differential equations of the form (\ref{eq:8}). Indeed, the 
underlying noise type need not take the form of a Brownian motion:  
depending on the context of the experiment the noise type may well be 
modelled more appropriately from a considerably broader family of 
processes known as L\'evy processes, of which Brownian motion is just 
one example. However, for each noise type (e.g., a Poisson process, a 
gamma process, a variance gamma process, and so on) there is a 
canonical way of formulating the signal detection problem \cite{BHY}, 
and deduce the corresponding 
conditional density process $\{\pi_{it}\}$. Then by defining $\phi_{it}=
\sqrt{\pi_{it}}$ and substituting the result in (\ref{eq:9}) we obtain the 
corresponding stochastic unravelling of the same Lindblad equation. 
Hence in each case we are left with the problem of signal processing 
in one form or another. It is also worth noting that the 
open quantum dynamics for the density matrix may take a form more 
general than the familiar Lindblad equation. For example, if the Markovian 
approximation used to deduce the dynamics of the reduced density matrix 
is not applicable, then one is lead to a corresponding stochastic unravelling 
for the wave function evolution that is manifestly non-Markovian \cite{DGS}. 
Whether these more general equations admit an 
explicit signal detection interpretation remains an open question, though 
we would speculate from the structure of the unravelling equation \cite{DGS} 
that the answer will be affirmative. 

At any rate, the foregoing analysis demonstrates that the stochastic 
unravelling of the 
Lindblad equation \textit{necessarily takes the form of the optimal signal 
detection}. In the Brownian case 
the dynamical equation (\ref{eq:10}) that gives rise to a seemingly random 
and yet purposeful evolution is a Hilbert space formulation 
of the Kushner equation \cite{Kushner,Bucy} 
in signal processing. Importantly, the Wiener process 
$\{W_t\}$ appearing in (\ref{eq:10}) does not represent noise, but is the 
innovations process. Hence the evolution of the wave function 
follows the random path that is determined by the ``best'' estimate of 
the underlying signalling problem: It is not the case that the state 
of an open quantum system is randomly perturbed by means of a noisy 
Brownian motion (as in the classical Brownian particle). For sure the state 
is influenced by the noise $\{B_t\}$, but its evolution follows the path 
determined only by the arrival of new information resulting from the 
interaction with the environment regarding the stable state of the system, 
corresponding to the eigenstates of the Lindblad operator. 
In other words, the wave function of the system is the representation of 
the conditions of their uncertain environments. 

\section*{Quantum dynamics for motion tracking}

This feature of quantum dynamics resembles, at least at an intuitive level, 
the behaviour of biological systems. With this in mind we attempt to model 
motions of biological systems -- plants in this case -- in 
response to changing environments using stochastic Schr\"odinger equations. 
To characterise the tracking motion we consider a model of the type 
discussed, for instance, by Diosi \cite{Diosi1}, in which the Lindblad 
operator is the 
position operator: ${\hat L}=\frac{1}{2}\sigma{\hat Q}$, where $\sigma$ is a 
constant. The effect of the Lindbladean part of the dynamics is to 
localise the wave function at the hidden ``true'' value of 
the position, which in the present context may represent the location of the 
light source. Plants do not possess sophisticated visual sensors, so 
\textit{a priori} the position of the light source is hidden to them. However, 
they possess a range of photoreceptors 
to detect and analyse incident light so as to regulate their responses to the 
environments \cite{Huq}. From the viewpoint of signal detection, the target 
-- e.g., the position of the sun -- is moving, and this will be modelled by 
choosing the Hamiltonian to be just the momentum operator: ${\hat H}=
\mu{\hat P}$, where $\mu$ is the rate parameter. Thus we have the 
dynamical equation 
\begin{eqnarray} 
\rd \psi_t(x) = -\ri \mu {\hat P} \, \psi_t(x)\, \rd t  + 
\textstyle{\frac{1}{2}} \sigma \big( {\hat Q} - 
\langle {\hat Q}\rangle_t \big) \psi_t(x) \, \rd W_t - \textstyle{\frac{1}{8}} 
\sigma^2 \big( {\hat Q} - \langle {\hat Q}\rangle_t \big)^2 \psi_t(x) \, \rd t .
\label{eq:11} 
\end{eqnarray}
The idea that we propose here therefore is to regard the squared wave 
function $\pi_t(x)=|\psi_t(x)|^2$, where $\psi_t(x)$ is the solution to 
(\ref{eq:11}), as representing the probability distribution of the location of 
the light source, as ``perceived'' by the plant (along the east-to-west straight 
line, which we take to approximate the actual motion of the sun on the 
upper hemisphere via vertical projection; 
the model here is understood to capture the qualitative behaviour of motion 
tracking for illustrative purposes). In other words, the integral 
$\int x\,\pi_t(x)\,\rd x$ 
represents the best estimate of the location of the light source arrived at 
by the plant from the data gathered through a range of receptors. 

For simplicity let us assume that the initial state is a standard Gaussian 
state: $\psi_0(x) = (2\pi)^{-1/4} \exp(-\frac{1}{4}x^2)$. The signal detection 
solution to (\ref{eq:11}) is then as follows. We let $X$ be a standard normal 
random variable with mean zero and variance one, and consider the ``signal'' 
process $X_t = X + \mu t$. The observation of the signal, however, is 
obscured by a Brownian noise, giving rise to the information process 
\begin{eqnarray} 
\xi_t = \sigma \int_0^t X_s \, \rd s + B_t.
\end{eqnarray} 
Then by considering the conditional density 
${\mathbb P}(X_t=x|\{\xi_s\}_{s\leq t})$ that determines the best estimate 
for the location of the target 
we deduce, after a short calculation, that the full solution to (\ref{eq:11}) -- 
not merely the asymptotic steady-state solution \cite{GRW,Diosi2} -- with the 
Gaussian initial state is given by 
\begin{eqnarray} 
\psi_t(x) = \left( \frac{1+\sigma^2t}{2\pi}\right)^{\frac{1}{4}}
\exp\left( -\frac{\left[ (1+\sigma^2t)x - \sigma\xi_t - \mu t (1+\frac{1}{2}
\sigma^2 t ) \right]^2 }{4(1+\sigma^2t)} \right) ,
\end{eqnarray} 
where the innovations Brownian motion $\{W_t\}$ in (\ref{eq:11}) 
is related to the information process $\{\xi_t\}$ according to 
\begin{eqnarray}
\rd W_t = \rd \xi_t - \sigma \left( \frac{\sigma\xi_t + \mu t (1+\frac{1}{2}
\sigma^2 t)}{1+\sigma^2t} \right) \rd t . 
\end{eqnarray} 

In the present case where the initial state is Gaussian, the best estimate 
for the position of the light source:
\begin{eqnarray} 
\int_{-\infty}^\infty x\,\pi_t(x)\,\rd x = \frac{\sigma\xi_t + \mu t (1+\frac{1}{2}
\sigma^2 t)}{1+\sigma^2t} 
\label{eq:x22} 
\end{eqnarray} 
is a linear function of the information process $\{\xi_t\}$, which seems 
reasonable. More generally, in the case of an arbitrary initial state 
$\psi_0(x)$, writing $p(x)=|\psi_0(x)|^2$, letting $X$ be a random variable 
with the density $p(x)$, and setting $X_t=X+\mu t$, the solution to 
(\ref{eq:11}) is determined by the conditional probability process: 
\begin{eqnarray}
\pi_t(x) = \frac{p(x) \, \exp\left( \sigma(x-\mu t)\left(\xi_t-\frac{1}{2}\sigma 
xt\right)\right)}{\int p(x) \, \exp\left( \sigma(x-\mu t)\left(\xi_t-\frac{1}{2}\sigma 
xt\right)\right) \, {\rm d}x} . 
\end{eqnarray}
That is, $\psi_t(x)=\re^{{\rm i}\theta(x)}\sqrt{\pi_t(x)}$, where $\theta(x)$ 
in the initial phase. 
The relatively simple model constructed here appears 
effective in characterising the simple solar-tracking motion whereby the 
plant orients towards the direction of the sun but with small errors. In 
particular, if the distribution of the error in the orientation of the plant 
leaves and flowers were normally distributed, then we find that the 
Gaussian model (\ref{eq:x22}) would be appropriate.

\section*{Quantifying the information extracted}

We are interested in the question on how much information extraction is 
needed so as to deduce the location of the unknown moving target $\{X_t\}$. 
In a spontaneous localisation model like (\ref{eq:11}) the large-time behaviour 
of the system is not physically relevant, for, in the limit $t\to\infty$ the wave 
function converges to (the square root of) a delta function at the point corresponding to 
the true value of $X$, but this requires an 
infinite amount of information, 
or, equivalently, infinite reduction in entropy. 

In the case of solar tracking of plants, on the other, the sun is not a point 
particle. Likewise, the leaves or flowers are not positioned perfectly 
perpendicular to the incident light. Hence the localisation 
model (\ref{eq:11}) is only relevant up to the time when a good progress is 
made in terms of tracking the moving sun. This intuitive idea can 
be made precise by studying the uncertainty (variance) measure of $X$: A 
good progress is made if the conditional variance of $X$ is reduced to a 
fraction of the initial variance \cite{BH2,LH,Adler}. 
In the present context, 
this timescale $\tau$ is given by $\tau\sim1/\sigma^2 \Delta X^2$, 
where $\Delta X^2$ is the initial variance of $X$. 
 
To quantify the information extraction, therefore, we consider the reduction 
$S_0-S_\tau$ of the Shannon-Wiener entropy 
\begin{eqnarray} 
S_t = -\int \pi_t(x) \log \pi_t(x) \rd x, 
\end{eqnarray}  
where $\pi_t(x)=|\psi_t(x)|^2$. 
Note that we are not concerned with the von Neumann entropy 
here, which represents the observer's knowledge, or lack of it, of the system; 
whereas the Shannon-Wiener entropy represents the system's lack of 
knowledge of the environment. To see this, we recall that density matrix of 
the system is given by 
\begin{eqnarray}
\rho_t(x,y) = {\mathbb E} \left[ {\overline{\psi_t(x)}} \psi_t(y)\right],
\end{eqnarray} 
where ${\mathbb E}$ denotes expectation over all random paths for the 
information process $\{\xi_t\}$, and it is $\rho_t(x,y)$ that fulfils the 
deterministic Lindbladean dynamical equation describing the time 
evolution of the open system. Because an (intelligent) observer 
is unaware which paths the information process has chosen, but knows 
only of its statistical distribution, the ``state'' of the system as perceived 
by the external observer is given by the density matrix $\rho_t(x,y)$. 
The von Neumann entropy 
\begin{eqnarray} 
-{\rm tr}(\rho\ln\rho) = -\iint \rho_t(x,y)\ln\rho_t(y,x)\, \rd y\, \rd x 
\end{eqnarray} 
thus represents the observer's 
lack of knowledge of the exact state of the system. In contrast, since 
$\pi_t(x)=|\psi_t(x)|^2$ represents the (unintelligent) system's knowledge 
of the environment, gathered by optimally processing the noisy 
information about the environment, it is the Shannon-Wiener entropy 
that represents the system's lack of knowledge of the environment.

Because the entropy process 
$\{S_t\}$ is stochastic, and depends on the information process $\{\xi_t\}$, it 
can increase or decrease, but on average it decreases \cite{BH2}. 
Hence we are interested in the averaged entropy reduction 
$\Delta S =S_0-{\mathbb E}[S_\tau]$ for the information gain. 
In the special case 
where the initial state is Gaussian, however, the associated entropy is 
deterministic and is given by 
\begin{eqnarray} 
S_t = \frac{1}{2}\left[1+\log(2\pi) - \log(1+\sigma^2t)\right],
\end{eqnarray}  
from which it follows that $\Delta S = \frac{1}{2}\log(1+\sigma^2\tau)$. 
On the other hand the initial uncertainty of $X$ in this example is unity 
so that $\tau=\sigma^{-2}$. Hence $\Delta S = 
\frac{1}{2}\log 2$ and we find that the 
amount of information processed to track the motion is half of 
one bit. Allowing for a variability in the initial wave function, we thus conclude 
that the amount of information processed to track the motion is at most of 
order few bits.

\section*{Information erasure and heat production}

It seems reasonable to assume that much of the information processed for 
plant movements are not stored in the plant indefinitely. (Note that this need 
not apply universally. For instance, there are suggestions that Mimosa plants 
can be trained to learn certain behaviours, which it can remember for more 
than a month \cite{GRDM}.) But the loss of processed information has to be 
distinguished from information that are encoded in their genes. Take, 
for instance, the seed of a plant buried in dark soil. Information 
encoded in the seed cells sends roots towards the direction of the gravitational 
pull, and stems in the opposite direction -- the so-called gravitropism. 
However, plants \textit{a priori} do 
not possess information about their environments, so 
measurements are performed to detect the direction of the 
gravitational field. In particular, if the direction of the gravitational force changes, 
then roots will change the orientation of their growths to 
accommodate this change of the environment \cite{Knight}. 

We are not concerned here with the actual mechanism by which plans 
detect gravitational field, which has only been unravelled relatively recently 
\cite{Baluska,Kiss}. What concerns us is the information-processing aspect 
of the growth orientation selection, and this can be modelled heuristically 
using a stochastic Lindbladean dynamical equation for a two-level system 
with ${\hat H},\,{\hat L}\propto{\hat\sigma}_z$; the solution to which in the 
quantum context are known \cite{BH1,BH2}. Specifically, we can think of 
a uniform magnetic field in the vertical $z$-direction surrounding a 
spin-$\frac{1}{2}$ particle in quantum mechanics as the analogue of the 
gravitational field surrounding the seed in biology. 
The direction of the spin then represents the orientation 
of the root growth. 
Under the dynamics governed by such stochastic Schr\"odinger 
equation the spin of the particle, which initially could be pointing in any 
orientation, will follow a random path such that eventually it will line up 
parallel to  
the direction of the field, and this reorientation of spin can be interpreted as 
modelling the root reorientation in gravitropism. 
Thus in this model the information process is 
$\xi_t=\sigma L t + B_t$, where the random variable $L$ takes the value 
$+1$ with probability $p=|\langle\uparrow\!|\psi_0\rangle|^2$, and takes the 
value $-1$ with probability $1-p=|\langle\downarrow\!|\psi_0\rangle|^2$. 
Here, $|\psi_0\rangle$ is the initial wave function of the system, and 
$(|\!\uparrow\rangle,|\!\downarrow\rangle)$ 
are the two eigenstates of the Pauli matrix ${\hat\sigma}_z$, corresponding 
to the spin-up and spin-down states. The effect 
of the Hamiltonian ${\hat H}=g{\hat\sigma}_z$ is then to generate the 
nutation of the root with angular frequency determined by the parameter 
$g$, while the effect of the Lindblad operator ${\hat L}=\gamma 
{\hat\sigma}_z$ is to generate gravitropism at a rate that is governed by 
the parameter $\gamma$. 

Because the problem here is of binary nature, of order one bit of 
information is processed to decide the growth orientation. This 
information is transient; 
measurements are performed repeatedly (by continuous monitoring) to 
reaffirm the orientation of the gravitational field \cite{Sato}. 
Thus, one bit of information must be regularly erased from each 
information-bearing unit inside the cells; most likely in the elongation 
zone cells, for, it seems implausible that such information 
is transmitted and stored elsewhere in the plant, because this will require 
additional resource for error corrections. 

Returning to heliotropism, take, for instance, the case of 
sunflowers. While mature sunflowers are permanently facing east, young 
sunflowers trace the motion of the sun. After sunset, the flowers then turn 
around from west to east and await the sunrise. This indicates a memory 
effect. Indeed, if one were to rotate a pot of a young sunflower plant by, say, 
$\pi/2$, then the sunflower now turns back and forth between north 
and south for a few days before stopping \cite{Harmer2}. In other words, 
the memory is lost only after a while. According to our rough estimate above, 
this results in the erasure of at most few bits of information. On the other 
hand, in the case of a daisy flower, overnight the flower tend to orient in a 
random direction, indicating a faster loss of information. 

It is known that information erasure is necessarily 
accompanied by energy consumption and hence heat production, resulting 
in an increase of environmental entropy \cite{LR}. In particular, the minimum 
amount of energy consumption required for the erasure of one bit of 
information at temperature $T$ is $k_{\rm B}T\log 2$. Based on 
our estimate, a daisy flower at summer-night temperature 
(say, 290 K) will thus consume at least of order $10^{-2}$~eV of energy per 
each information-encoding unit. This 
in turn results in heat production. A detection of this effect is likely to be 
difficult, for, the magnitude of energy involved for information erasure 
based on our estimate is likely to be rather small compared to that for 
plant movements. 
Further, the environment of plants are far removed from the highly 
controlled environment in which such an effect has been detected in a 
quantum system \cite{Berut}. 

The precise mechanism of how plants store information is not fully 
understood. Hence it is not known how many information-storing 
units are contained in a given plant, making it difficult to estimate the total 
energy cost of information erasure. To get a better 
intuition for the scale involved, therefore, let us consider, 
as an example, the circumnutation of 
a common bean plant \cite{Calvo} by assuming that each cell contains at 
most a single such unit. For the bean plant to identify the location of an object 
like a pole to climb up within a $\pm10^\circ$ angular window with $95\%$ 
confidence, say in a recent experimental setup \cite{Calvo}, 
it must process between 15 and 20 bits of information. This estimate 
follows from the assumption that the \textit{a priori} probability of finding 
an object is uniform over the circle around the plant. Assuming that the 
\textit{a posteriori} distribution can be described by any one of the standard 
circular distributions such as the von Mises or the wrapped normal distribution, 
one arrives at this estimate. Now if the 
object is removed, this information must be erased. If the plant can process 
information relatively close to the Landauer limit, say, $10^4$ times the 
limit, then the cost of erasure is of order $10^{4}$~eV, which would be 
approximately just 
below $1\%$ of the total energy consumed by the cell \cite{Milo}. Intuitively this 
is a plausible figure, given the limited energy resource available to biological 
systems. Or, putting the matter differently, if information is processed 
significantly less efficiently, say at the level of our everyday computers, then 
the erasure cost becomes too high for survival. Our hypothesis 
that biological systems can process information close to the Landauer limit is 
consistent with the empirical observation that plants produce very little heat 
(except for thermogenic plants that purposefully produce heat 
\cite{Robinson}).

\section*{Second law in biology}

In accordance with the Landauer principle the erasure of information will 
increase the entropy of the environment. It is tempting therefore to conjecture 
that biological systems operate by extracting information from their 
environments, processing them, and arriving at the best estimate of the 
state of the environment for the purpose of adaptation. 
This will result in the gain of information, and thus 
reduction in entropy. However, some, or much of the processed information 
is lost, resulting in increasing entropy. The process of information erasure 
then must lie at the heart of ageing -- or arrow of time -- in biological systems. 

It should be added that while the present paper concerns the dynamical 
mechanism of the information-processing of the environmental conditions of 
biological systems, our estimates show that the erasure cost, and hence 
entropy production, of genetic information encoded 
\textit{a priori} in biological systems would be substantially more significant 
than that of the \textit{a posteriori} processed information. This is consistent 
with the empirical fact that the life span of a biological system on average 
reduces when genetic information encoded in the system is lost. Hence in 
terms of biological arrow of time, we argue that the erasure of processed 
information will not constitute the dominant contributing factor.

\section*{Model calibration}

\begin{figure}[h]
      \centering
       \subfloat[$\gamma=0.25$: slow reorientation]
       {\includegraphics[width=0.45\textwidth]{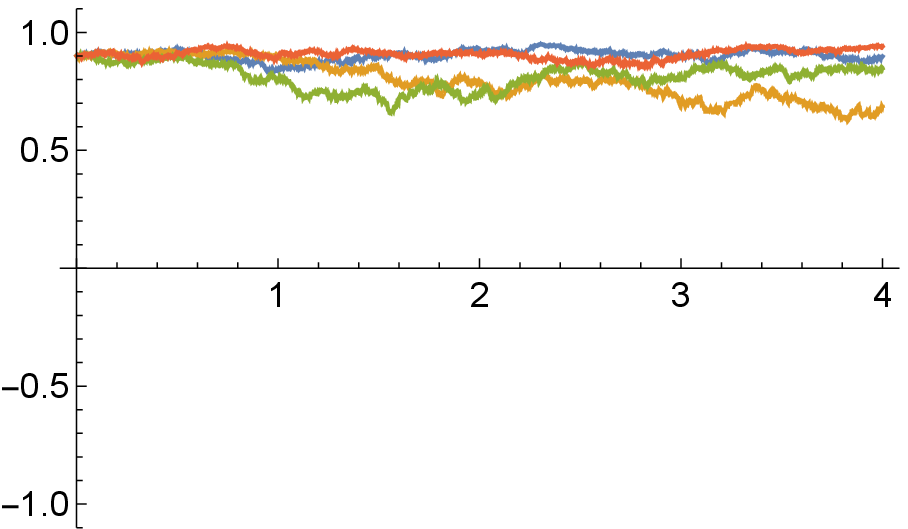}}\hfill
       \subfloat[$\gamma=0.95$: fast reorientation]
       {\includegraphics[width=0.45\textwidth]{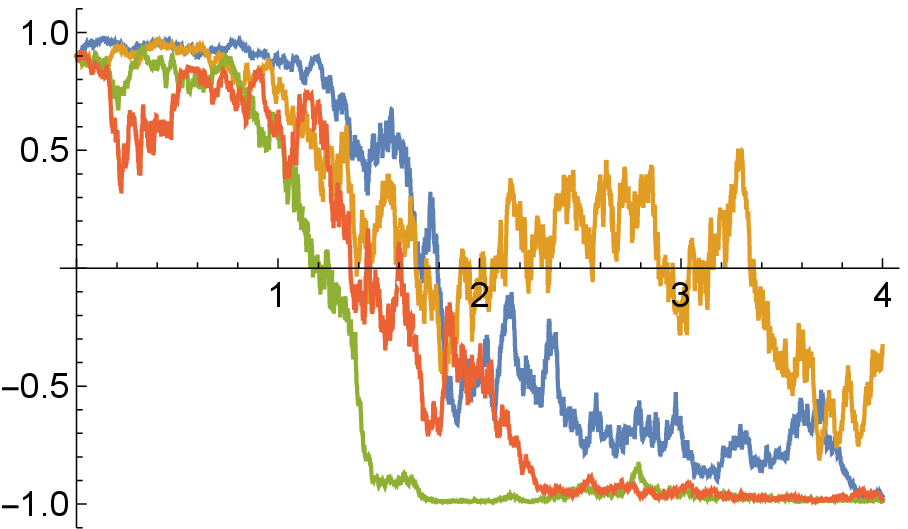}}\hfill      
\caption{\textit{Model simulation of the binary root reorientation}. In a model with 
${\hat L}=\gamma{\hat\sigma}_z$, assume that the plant is initially 95\% 
confident that the gravitational force is pointing down (the signal $L=-1$); but 
at time zero it is flipped upside-down (the signal $L=+1$). Depending on the 
value of the parameter $\gamma$ the 
reorientation (and hence the loss of the initial memory) can be slow or fast. 
Four sample paths for the order parameter $\langle {\hat\sigma}_z\rangle_t$ 
are shown here each for the two chosen values of $\gamma$. 
The order parameter represents the vertical component 
of the unit directional vector of the root. Because the reorientation timescale 
is proportional to $\gamma^{-2}$, although in all cases the 
root will eventually point down, this process can take a long time if $\gamma$ 
is small (the left panel); whereas for a larger value of $\gamma$ the 
reorientation takes place rapidly (the right panel).}
\label{fig1} 
\end{figure}

How can model parameters be estimated against data? In our simple tracking 
model we have the rate parameter $\sigma$, which determines the timescale 
that plants can orient towards the light source. For instance, leaves of Cornish 
mallow can reorient as rapidly as $2\pi/9$ per hour \cite{Koller}, indicating 
that $\sigma$ takes a large value. (Recall that the response 
timescale is inverse proportional to $\sigma^2$ so that a rapid response 
would imply a large $\sigma$.) If plants exhibit a memory effect, like 
circadian rhythm of sunflowers \cite{Harmer2}, then either the 
variance of the signal is small or else the value of $\sigma$ is 
small. In the case of gravitropism, using the model described above, if we let 
${\hat L}=\gamma{\hat\sigma}_z$, 
then the parameter $\gamma$ can be calibrated from studying the timescale 
of root reorientation by turning the growing plant still in the soil upside-down. 
Figure~\ref{fig1} shows how the response time of the plant can be slow or 
fast, depending on the value of $\gamma$. 

\section*{Discussion}

In summary, we have shown how stochastic unravelling of the Lindblad 
equation for open system dynamics in quantum mechanics necessarily 
takes the form of the equation for the best information processing in signal 
detection. This observation suggests that the notion of an 
optimal information-processing 
capability is fundamental to the laws of nature at the quantum level. Such a 
statement may at first seem counterintuitive, for, one tends to associate 
some form of intelligence to the concept of information processing. In physics 
one is more accustomed to the idea of a variational principle to arrive at 
laws of nature. However, our conclusion can in fact be reached by means of 
a variational argument, albeit in noisy environments. Namely, the dynamical 
equation for the state of the system can be derived by demanding that the 
average entropy reduction of the system is maximised. 
Indeed, the entropy and variance -- the two uncertainty 
measures -- are closely related in the context of signal processing. 
This follows from 
the fact \cite{BH2} that the Shannon-Wiener entropy associated 
with the stochastic Schr\"odinger equation (\ref{eq:8}) satisfies the relation 
\begin{eqnarray}
{\mathbb E}\left[ S_\tau \right] = \half \sigma^2 \int_\tau^\infty 
{\mathbb E}\big[\Delta L_t^2\big]\, \rd t , 
\end{eqnarray}
where $\Delta L_t^2=\langle\psi_t| ({\hat L}-\langle{\hat L}\rangle_t)^2
|\psi_t\rangle$ 
is the conditional variance of the Lindblad operator. Hence minimisation of 
the quadratic error is linked to maximisation of the entropy reduction.  

Given that biological systems like plants use sophisticated mechanisms 
to process external signals concerning light, water, temperature, gravity, 
organic compounds, and 
so on, and adapt their behaviours accordingly, we postulated that the stochastic 
unravelling of the Lindblad equation in open quantum systems can be 
applied to model the dynamical behaviours of plants in open environments. 
We considered a localisation model to capture 
qualitatively the tracking motion, to arrive at a rough 
estimate of the quantity of information processed. The idea that information 
processing in biological systems must be viewed as fundamental has been 
advocated before \cite{Quadtler,Binder,Mescher}, 
but here we have been able to 
quantify this in a dynamical context. Empirical observations suggest that 
some of the processed information is erased, from which we postulated 
(a) that biological systems must operate relatively close to the Landauer 
limit of computation; and (b) that information erasure must underlie the ageing 
of biological systems in a fundamental way. 

It is worth remarking that the specific models of stochastic Schr\"odinger 
equations considered above for the characterisation of plant motions can 
be presented by use of purely classical signal detection techniques. While 
this is consistent with our thesis that signal processing capability is 
fundamental to laws of nature including quantum theory, it also implies 
that the resulting estimates in the context of biology, 
at least in the examples considered here, in principle 
could have been obtained  
purely in terms of classical signal detection, without referencing 
quantum theory. 
That said, it would be technically more challenging to pose 
a model for the spherical nutation of roots using a purely classical language, 
whereas our quantum spin-$\frac{1}{2}$ model offers a highly effective and 
simple treatment of the matter. More importantly,  
we believe that in many realistic setups in 
biology, more adequate descriptions can be obtained by means of more 
quantum-mechanical models for which the commutator of the Hamiltonian 
${\hat H}$ and the Lindblad operator ${\hat L}$ is nontrivial (that is, 
$[{\hat H},{\hat L}]$ neither vanishes nor is proportional to the identity). In 
the biological context, the Lindblad operator represents adaptation, while 
the Hamiltonian represents changes of the environment. (For example, in 
the simple motion-tracking model considered here, the Lindblad operator 
has the effect of orienting towards the location of the sun, 
whereas the Hamiltonian changes the location of the sun.) It is inadvertently 
the case in biology that adaptation is possible so long as environmental 
changes are sufficiently slow, whereas a fast change can be catastrophic to 
the survival of the 
biological system, leading to a very different dynamical behaviour. This is 
already seen even in the simple phenomenon of heliotropism \cite{Harmer1}. 
Such a transition can be described by a stochastic 
Schr\"odinger equation for which $[{\hat H},{\hat L}]\neq0$ in a nontrivial 
way, because in such a model one typically encounters a phase transition in 
the dynamical behaviour of the system \cite{Bassi,BL}, depending on the 
relative strengths of ${\hat H}$ and ${\hat L}$. 
Conversely, without the lack of commutativity it is not 
possible to describe such critical phenomena seen in biology and ecology; 
yet, the notion of incompatible observables is one of the signatures of 
quantum theory. We argue that this observation 
justifies our unified approach to model dynamical behaviours of quantum 
and biological systems in open environments, and hope to develop more 
general theories elsewhere. 

We conclude by remarking that there is an ongoing debate in plant science 
concerning the notion of consciousness and sentience 
in plants \cite{Reber,Taiz,Calvo2,Calvo3}. This follows from advances in 
detecting 
plants' remarkable abilities in observing, analysing, and adapting to the 
changing environments that surround them. Our postulate that 
information-processing capability is part of the laws of nature, thus not 
requiring any intelligence, might shed a new light on this debate.

\section*{Acknowledgements}

The author thanks L. P. Hughston, 
B. K. Meister, E. M. Graefe, G. Piskov, P. Calvo, F. Balu\v{s}ka, 
J. Al-Khalili, A. Rocco, S. Saunders, and A. Trewavas for stimulating 
discussions; and acknowledges support from the Russian Science 
Foundation (grant 20-11-20226), the London Mathematical 
Society (grant URB-2021-43), and the John Templeton Foundation 
(grant 62210). The opinions expressed in this publication are those of 
the author and do not necessarily reflect the views of the 
John Templeton Foundation.

\end{document}